\documentclass[10pt]{article}

\usepackage{amsmath}
\usepackage{amssymb}  
\usepackage[T1]{fontenc}
\usepackage{pifont}
\usepackage{pstricks}
\usepackage{amsfonts}
\usepackage{graphicx}
\usepackage{amsmath}
\usepackage{authblk}
\usepackage{pstricks} 
\usepackage{pstricks-add}
\usepackage{caption}
\usepackage{placeins}
\usepackage{pstricks}
\usepackage{pdfpages}
\usepackage{framed}
\usepackage{bigints}
\usepackage{cancel}
\usepackage{bm}
\usepackage{epstopdf}
\usepackage{tikz}

\def\rr2dot{\mathop{\bf r}\limits}
\def\x2dot{\mathop{x}\limits}
\def\y2dot{\mathop{y}\limits}

\def\bfy2dot{\mathop{\bf y}\limits}
\def\z2dot{\mathop{z}\limits}
\def\min{\mathop{min}\limits}

\def\csi2dot{\mathop{\xi}\limits}
\def\et2dot{\mathop{\eta}\limits}
\def\bet2dot{\mathop{\beta}\limits}
\def\t2dot{\mathop{\theta}\limits}
\def\s2dot{\mathop{\sigma}\limits}
\def\d2dot{\mathop{\delta }\limits}
\def\q2dot{\mathop{q}\limits}
\def\l2dot{\mathop{\lambda}\limits}
\def\ps2dot{\mathop{{\cal E}}\limits}
\def\tet2dot{\mathop{\theta}\limits}
\def\bfx2dot{\mathop{\bf x}\limits}
\def\bfy2dot{\mathop{\bf y}\limits}
\def\bfq2dot{\mathop{\bf q}\limits}
\def\bfr2dot{\mathop{\bf r}\limits}
\def\bbfq2dot{\mathop{\bar {\bf q}}\limits}
\def\w2{\mathop{W}\limits}
\def\xgrande2dot{\mathop{\bf X}\limits}
\def\p02dot{\mathop{P}\limits}
\def\a2dot{\mathop{A}\limits}

\newtheorem{lem}{Lemma}
\newtheorem{prop}{Proposition}
\newtheorem{cor}{Corollary}

\newtheorem{rem}{Remark}
\newtheorem{exe}{Example}

\newtheorem{proof}{Proof}

\DeclareMathOperator{\rank}{rank}

\title{On the invertibility of saddle-point matrices: exact algebraic conditions and applications to constrained dynamics}

\author{F.~Talamucci}
\affil{{\it DIMAI, Dipartimento di Matematica e Informatica ``Ulisse Dini''},\\
{\it	Universit\`a degli Studi di Firenze, Italy}\\
{\it	e-mail: federico.talamucci@unifi.it}}
\date{}

\begin{document}
		
	\bibliographystyle{plain}
	
	\setcounter{equation}{0}

	\maketitle
	
	\vspace{.5truecm}
	
	
	\vspace{.5truecm}
	

	\vspace{.5truecm}
	

\begin{abstract}
The solvability of constrained dynamical systems governed by D'Alembert--Lagrange mechanics relies directly on the invertibility of the coefficient matrix describing the underlying saddle-point structure, which couples system inertia with kinematic constraints. Standard formulations often rely on overly restrictive sufficient assumptions, such as global positive definiteness or semi-definiteness of the mass matrix, frequently combined with generalized inverses to handle singular mass distributions. In this work, we establish the purely necessary and sufficient algebraic conditions for matrix invertibility: the kinematic independence of the constraints and a kernel intersection condition between the allowed directions of motion and the induced inertial forces. We provide a physical interpretation of this requirement, demonstrating that it eliminates non-zero admissible motions whose generated inertial forces are entirely absorbed by constraint reactions. Furthermore, by completely avoiding generalized inverses, this formulation offers a unified and coordinate-free framework capable of handling massless components and indefinite inertia structures within a computationally robust setting.
\end{abstract}

\section{Introduction}

\noindent
A KKT system (standing for Karush-Kuhn-Tucker, \cite{karush1939, kuhn1951}) is a block linear system with a specific saddle-point structure that arises in many areas of applied mathematics, particularly in constrained optimization, multibody robotics, mechanics of constrained systems, continuum mechanics (e.g., incompressible fluids), biomechanics, and finite element methods.
For a comprehensive analysis of KKT optimality conditions and numerical optimization, classic reference texts include Nocedal and Wright \cite{nocedal2006} and Boyd and Vandenberghe \cite{boyd2004}. Furthermore, from a numerical linear algebra perspective, the seminal survey by Benzi, Golub, and Liesen \cite{benzi2005} (and the literature cited therein) serves as a cornerstone reference for the algebraic structure, properties, and numerical solution of saddle-point problems.

\noindent
In this paper, the standard matrix form of the KKT system is given by
\begin{equation}
	\label{Lintr}
	\begin{pmatrix} {\Bbb O} & A \\ A^T & W \end{pmatrix} \begin{pmatrix} {\bm \lambda} \\ {\bf x} \end{pmatrix} = \begin{pmatrix} {\bm \beta} \\ {\bm \alpha} \end{pmatrix},
\end{equation}
where ${\Bbb O} \in \mathbb{R}^{m \times m}$ is the zero matrix, $A \in \mathbb{R}^{m \times n}$ with $m < n$, and $W \in \mathbb{R}^{n \times n}$, while ${\bf x}, {\bm \alpha} \in \mathbb{R}^{n \times 1}$ and ${\bm \lambda}, {\bm \beta} \in \mathbb{R}^{m \times 1}$ are column vectors. Accordingly, the KKT coefficient matrix is defined as $L = \begin{pmatrix} {\Bbb O} & A \\ A^T & W \end{pmatrix} \in \mathbb{R}^{(m+n) \times (m+n)}$.

\noindent
Although it is not possible to assign a single, unique interpretation to the matrices and vectors in (\ref{Lintr}) given the wide variety of applicable domains, in very general terms matrix $A$ plays the role of a ``constraint matrix'' via the restrictions $A{\bf x}={\bm \beta}$. Meanwhile, matrix $W$ captures the dynamics or core structure of the system under study (e.g., optimization, motion of holonomic or nonholonomic systems) through the equation $W{\bf x}+A^T{\bm \lambda}= {\bm \alpha}$. In particular, the multiplier vector ${\bm \lambda}$ reveals how the system's evolution is modified by the imposed constraints relative to the unconstrained setting (${\bm \lambda} = {\bf 0}$).
For instance, systems of this form arise naturally as the first-order necessary optimality conditions for the quadratic problem (\cite{boyd2004, nocedal2006})
$$
\min \frac{1}{2}{\bf x}^TW{\bf x}-{\bm \alpha}^T{\bf x} \quad \text{subject to} \quad A{\bf x}={\bm \beta}.
$$
As extensively detailed in the literature on saddle-point problems \cite{benzi2005}, the primary hallmark of these block-structured matrices is their inherent indefiniteness: they balance the minimization of an energy functional with respect to the physical coordinates against the maximization enforced by the constraint multipliers.

\noindent
A well-posed physical system requires the unique determination of both ${\bm \lambda}$ and ${\bf x}$, for any set of data ${\bm \alpha}$ and ${\bm \beta}$. This uniqueness hinges upon the invertibility of the coefficient matrix $L$.

\noindent
The focus of this work is precisely the mathematical investigation into the invertibility of $L$, establishing a necessary and sufficient criterion to this end. Although simple and admitting a straightforward geometric interpretation, to the best of our knowledge, the derived conditions do not appear directly in this formal setting in the existing literature.
The proposed algebraic requirement is further explored in relation to specific properties of the system matrices (e.g., symmetry, invertibility, definiteness). Simultaneously, a comparison is drawn with existing criteria (such as positive definiteness of $W$, LBB ellipticity, and the Udwadia--Phohomsiri assumption), encompassing both purely sufficient conditions and necessary and sufficient ones.

\noindent
From a modeling perspective, our main focus is directed towards the case where system (\ref{Lintr}) governs the motion of a mechanical system subjected to constraints. In classical analytical mechanics, $W$ corresponds to the system's ``generalized'' mass matrix $M$. Consequently, standard literature regularly invokes the assumption of strict positive definiteness \cite{lemos}, under which $L$ is guaranteed to be non-singular provided $A$ has full row rank. In modern computational mechanics, optimization-based control, and systems with auxiliary coordinates or massless components, $W$ is frequently only semi-definite or even indefinite. To deal with singular mass matrices, classical frameworks - most notably the fundamental equation of constrained motion by Udwadia and Phohomsiri \cite{udwadia2006} - rely heavily on the Moore-Penrose pseudoinverse. While mathematically elegant, pseudoinverse-based formulations introduce potential numerical instabilities near rank changes and demand high computational overhead for large-scale systems.
Alternatively, in the context of finite element analysis and constrained optimization, existence and uniqueness are often framed around the LBB inf-sup condition \cite{brezzi1974}, which enforces strict positivity of $W$ restricted to the constraint kernel ($\ker (A)$). While weaker than global positive definiteness, the LBB condition remains a sufficient requirement rather than an exact equivalence.

\noindent
In addition to proposing a simple and ``economical'' invertibility criterion for $L$, which naturally encompasses existing criteria, we also extend the validity of D'Alembert--Lagrange dynamics to systems with indefinite or highly degenerate inertia structures. Furthermore, we provide a direct physical interpretation of the derived conditions, showing that they mathematically eliminate ``ghost'' motions - admissible virtual displacements whose resulting inertial forces are completely orthogonal to the subspace of allowed motion and thus physically unobservable.

\subsection{The physical model}

Before detailing the mathematical model, we emphasize that our focus on the saddle-point block structure (\ref{Lintr}) is specifically motivated by the constrained motion of mechanical systems. We consider systems subject to general kinematic constraints (whether holonomic or non-holonomic, linear or non-linear in the velocities) whose dynamics are governed by Lagrange's equations of the first kind incorporating constraint multipliers.

\noindent
Specifically, let us consider a mechanical system described by $n$ generalized coordinates $\mathbf{q} = (q_1, \dots, q_n)^T$ and subject to $m < n$ kinematic constraints 
\begin{equation}
	\label{vinc}
	\Phi_i(\mathbf{q}, \dot{\mathbf{q}}, t) = 0, \qquad i = 1, \dots, m,
\end{equation}
where $\dot{\mathbf{q}} = (\dot{q}_1, \dots, \dot{q}_n)^T$ denotes the vector of generalized velocities. The constraint matrix $A \in \mathbb{R}^{m \times n}$ is defined as the Jacobian with respect to the velocity variables:
\begin{equation}
	\label{a}
	A = \left( \dfrac{\partial \Phi_i}{\partial \dot{q}_j} \right)_{ \substack{i=1, \dots, m \\ j=1, \dots, n} } = \left( a_{i,j}(\mathbf{q}, \dot{\mathbf{q}}, t) \right)_{ \substack{i=1, \dots, m \\ j=1, \dots, n} }.
\end{equation}
We denote the rank of $A$ by $p \le m$. Unless specified otherwise, we assume throughout this work that the constraints are linearly independent, so that $A$ has full row rank, namely $p = m$.
\noindent
If $T({\bf q}, {\dot {\bf q}},t)$ is the kinetic energy of the system, the equations of motion can be written in the well known lagrangian form
\begin{equation}
\label{eqmoto}
\dfrac{d}{dt}\dfrac{\partial T}{\partial {\dot q}_j} - \dfrac{\partial T}{\partial q_j}=F_j +F_j^C, \qquad j=1, \dots, n
\end{equation}
where $F_j$ are the lagrangian components of the applied forces, which in the context of our study are assumed to be known functions of $({\bf q}, {\dot {\bf q}},t)$, and $F_j^C$ are the unknown  generalized constraint forces due to the presence of conditions (\ref{vinc}). 
In turn, $F_j$ can be split into applied forces $F_j^P({\bf q, {\dot {\bf q}}},t)$ derived from a generalized potential $U({\bf q}, {\dot {\bf q}}, t)$, namely $F_j^P = \frac{\partial U}{\partial q_j} - \frac{d}{dt}\frac{\partial U}{\partial \dot{q}_j}$ (which includes the standard case of a velocity-independent potential $U({\bf q}, t)$), and non--potential forces $F_j^N({\bf q, {\dot {\bf q}}},t)$ lacking this property. The equations of motion \eqref{eqmoto} take the form:
\begin{equation}
\label{eqmotol}
\frac{d}{dt}\frac{\partial \mathcal{L}}{\partial \dot{q}_j} - \frac{\partial \mathcal{L}}{\partial q_j} = 
F_j^N + F_j^C, \qquad j=1, \dots, n
\end{equation}
where $\mathcal{L}=T-V$, $V=-U$, is the Lagrangian of the system.

\noindent
One viewpoint - widely accepted in the literature and adopted here - transfers the unknown nature of the constraint forces to $m$ Lagrange multipliers $\lambda_1, \dots, \lambda_m$, according to the relation (sometimes also referred to as the Chetaev condition):
\begin{equation}
	\label{moltlagr}
	F_j^C = \sum\limits_{i=1}^m \lambda_i \dfrac{\partial \Phi_i}{\partial {\dot q}_j}, \qquad j=1, \dots, n,
\end{equation}
where $\lambda_i=\lambda_i({\bf q}, {\dot {\bf q}},t)$, $i=1, \dots, m$. For a more detailed discussion on formulating the equations of motion for systems subject to constraints of type (\ref{vinc}), we refer the reader to \cite{lurie}, \cite{neimark}, \cite{papast}.
Recalling (\ref{a}), the equations of motion can therefore be written as
\begin{equation}
	\label{eqmotolambda}
	\dfrac{d}{dt}\dfrac{\partial {\cal L}}{\partial {\dot q}_j} - 
	\dfrac{\partial {\cal L}}{\partial q_j} = F_j^N + \sum\limits_{i=1}^m a_{i,j}\lambda_i, \qquad j=1, \dots, n.
\end{equation}
In view of the subsequent topics, it is worth recalling the formalism of linear algebra in order to arrive at assumption \eqref{moltlagr}. This approach defines virtual displacements as the set of $n$-vectors ${\bf v}$ satisfying
\begin{equation}
	\label{spostvirt}
	A{\bf v}={\bf 0},
\end{equation}
where $A$ is the coefficient matrix \eqref{a}. For any fixed kinematic state $(\mathbf{q}, \dot{\mathbf{q}})$ and time $t$, the vectors ${\bf v}$ form the linear space $\ker(A)$ (the kernel of $A$) and are orthogonal to the $m$ row vectors of $A$, and hence to the space $\operatorname{Im}(A^T)$, where $\operatorname{Im}(M)$ denotes the column space of a matrix $M$. Since we assume that $A$ has full rank, $\dim(\operatorname{Im}(A^T))=m$ and $\dim(\ker(A))=n-m$, consistently with the relation $\ker(A)=(\operatorname{Im}(A^T))^\perp$. 
The ideal constraint hypothesis reads
\begin{equation}
	{\bf F}^C \cdot {\bf v}=0 \qquad \forall {\bf v}\in \ker(A),
\end{equation}
which implies ${\bf F}^C \in (\ker(A))^\perp=\operatorname{Im}(A^T)$. Therefore, ${\bf F}^C$ must be a linear combination of the rows of $A$, yielding relation \eqref{moltlagr}.

\noindent
The explicit computation of (\ref{eqmotolambda}) leads to the system 
\begin{equation}
\label{eqmotom}
\sum\limits_{r=1}^n M_{j,r}{\q2dot\limits^{..}}_r-\sum\limits_{i=1}^m a_{i,j}\lambda_i =\alpha_j, \qquad j=1, \dots, n   
\end{equation}
where 
\begin{equation}
\label{m}
M_{j,r} ({\bf q}, {\dot {\bf q}},t)=\frac{\partial^2 {\cal L}}{\partial {\dot q}_r \partial {\dot q}_j}
\end{equation}  
and 
$\alpha_j ({\bf q}, {\dot {\bf q}},t)=F_j^N+\frac{\partial {\cal L}}{\partial q_j}-\frac{\partial^2 {\cal L}}{\partial t \partial {\dot q}_j}-\sum\limits_{r=1}^n \frac{\partial^2 {\cal L}}{\partial q_r \partial {\dot q}_j}{\dot q}_r$.

\noindent
On the other hand, by differentiating with respect to $t$ the constraint equations (\ref{vinc}) one gets the set
\begin{equation}
\label{vincder}
\sum\limits_{j=1}^n a_{i,j}{\q2dot\limits^{..}}_j=\beta_i, \qquad i=1, \dots, m
\end{equation}
where $a_{i,j}({\bf q}, {\dot {\bf q}},t)$ are defined in (\ref{a}) and  $\beta_i ({\bf q}, {\dot {\bf q}},t)=-\sum\limits_{j=1}^n \frac{\partial \Phi_i}{\partial q_j}{\dot q}_j-\frac{\partial \Phi_i}{\partial t}$.

\begin{rem}
	Clearly, (\ref{vinc}) implies (\ref{vincder}); however, if we replace the latter conditions with the original constraints (\ref{vinc}), the initial compatibility conditions $\Phi_i(\mathbf{q}(0), \dot{\mathbf{q}}(0), 0) = 0$, $i = 1, \dots, m$ must be added to avoid an arbitrary constant in the reverse integration step from (\ref{vincder}) to (\ref{vinc}).
\end{rem}
 
\noindent
We can arrange the system of $n+m$ equations (\ref{eqmotom}) and (\ref{vincder}) into matrix--vector notation as follows:
 \begin{equation}
 	\label{eqmotosistlin}
 	L{\bf v}={\bf c}
 \end{equation}	
where:
 \begin{itemize}
\item 
$L({\bf q}, {\dot {\bf q}},t)$ is a square matrix of order $m+n$ structured in block form as
 	\begin{equation}
 		\label{L}
 		L=\left( \begin{array}{cc} \mathbb{O} & A \\ A^T & W \end{array} \right)
 	\end{equation}
with ${\mathbb O}$ denoting the $m \times m$ zero matrix, $W({\bf q}, {\dot {\bf q}},t) = (M_{i,j})_{i,j=1, \dots, n}$ being the $n \times n$ matrix defined in (\ref{m}), and $A({\bf q}, {\dot {\bf q}},t) = (a_{i,j})_{\substack{i=1,\dots,m \\ j=1,\dots,n}}$ is the $m \times n$ matrix defined in (\ref{a});
 	
\item
${\bf v}$ and ${\bf c}({\bf q}, {\dot {\bf q}},t)$ are column vectors of dimension $m+n$ defined by
 	\begin{equation}
 		{\bf v}=(-\lambda_1, \dots, -\lambda_m, \ddot{q}_1, \dots, \ddot{q}_n)^T, \quad
 		{\bf c}=(\beta_1, \dots, \beta_m, \alpha_1, \dots, \alpha_n)^T.
 	\end{equation}
 \end{itemize}

\noindent
In (\ref{L}), we recognize the structure of (\ref{Lintr}), up to the minus sign in front of the multipliers, which is immaterial.
Formulating the linear system (w.~r.~t.~$\lambda_i$ and $\ddot{q}_j$, $i=1, \dots, m$, $j=1, \dots,  m$) as in (\ref{eqmotosistlin}) serves a two-fold goal that we aim to investigate in this work: enabling the system to be cast into explicit normal form $\ddot{q}_j = \Psi_j({\bf q}, {\dot {\bf q}}, t)$ for $j=1, \dots, n$ (for suitable functions $\Psi_j$), and allowing the explicit determination of the multipliers $\lambda_i$, $i=1, \dots, m$, solely as functions of the variables $({\bf q}, {\dot {\bf q}}, t)$.

The invertibility of the matrix $L$ is the fundamental algebraic condition ensuring the existence and uniqueness of the solution to the dynamical system. Specifically, it guarantees the unique determination of the acceleration vector $\ddot{\mathbf{q}}$ and the Lagrange multipliers $\boldsymbol{\lambda}$ (representing the constraint reactions) given the applied forces.

\noindent
Attention is therefore focused on the invertibility: if $L$ is invertible, the equivalent relation ${\bf v}=L^{-1}{\bf c}$ allows both the multipliers and the accelerations to be explicitly expressed as functions of $({\bf q}, {\dot {\bf q}}, t)$.
The result presented here provides a necessary and sufficient criterion for the matrix $L$ to be invertible. We will then compare this result with other conditions (generally sufficient ones) available in the literature.
 
\section{The main result} 
 
\noindent 
We state the main result, which provides necessary and sufficient conditions for the matrix $L$ defined in (\ref{L}) to be invertible. Since the investigation is purely algebraic, we can consider generic blocks $A$ and $M$, not necessarily arising from the framework of the motion (\ref{m}) and (\ref{vincder}).

\subsection{Notations and preliminaries}

We briefly recall some standard notations and well-known results from linear algebra. 

\noindent
Given a matrix $B \in \mathbb{R}^{\mu \times \nu}$, with positive integers $\mu$ and $\nu$, 
we denote by $\ker(B)$ the kernel (or null space) of $B$, i.e., the vector space
$$
\ker(B) = \{ \mathbf{x} \in \mathbb{R}^\nu \mid B\mathbf{x} = \mathbf{0} \},
$$
where $\mathbf{0}$ is the zero vector of $\mathbb{R}^\mu$. The fundamental relation between 
the dimension of the kernel $\dim(\ker(B))$ (also referred to as the nullity of $B$) 
and the rank $\operatorname{rank}(B)$ is given by the rank-nullity Theorem:
$$
\operatorname{rank}(B) + \dim(\ker(B)) = \nu.
$$
Furthermore, we denote by $\operatorname{Im}(B)$ the image (or column space) of the matrix $B$, 
namely the vector space spanned by all linear combinations of the columns of $B$:
$$
\operatorname{Im}(B) = \{ \mathbf{y} \in \mathbb{R}^\mu \mid \mathbf{y} = B\mathbf{x} \text{ for some } \mathbf{x} \in \mathbb{R}^\nu \}.
$$
Accordingly, $\operatorname{Im}(B^T)$ represents the row space of $B$. It is a well-known fact that the dimension of the column space coincides with that of the row space and equals the rank of $B$:
$$
\dim(\operatorname{Im}(B)) = \dim(\operatorname{Im}(B^T)) = \operatorname{rank}(B).
$$
Following standard convention, $\operatorname{span}\{\cdot\}$ denotes the subspace spanned by the vectors within the curly brackets.
Finally, we reiterate a property (in fact, one already used previously) concerning orthogonal complements: 
denoting by $\mathbb{V}^\perp$ the orthogonal complement of a vector space $\mathbb{V}$ 
(i.e., the set of all vectors orthogonal to every vector in $\mathbb{V}$), the following identity holds:
\begin{equation}
	\label{imort}
	(\ker(B))^\perp = \operatorname{Im}(B^T),
\end{equation}
which yields the orthogonal decomposition $\mathbb{R}^\nu = \operatorname{Im}(B^T) \oplus \ker(B)$ 
(that is, every $\nu$-vector can be uniquely expressed as the sum of an element of $\operatorname{Im}(B^T)$ and an element of $\ker(B)$).

\subsection{Statement and proof of the main result}

\begin{prop}
	Let $L$ be a block matrix as defined in (\ref{L}). The matrix $L$ is invertible if and only if the following two conditions are simultaneously satisfied:
	\begin{itemize}
		\item[$1.$] The matrix $A$ has full rank $m$:
		\begin{equation}
			\label{cond1}
			\operatorname{rank}(A) = m.
		\end{equation}
		
		\item[$2.$] The only vector in the kernel of $A$ that is mapped by $W$ into an element of the image of $A^T$ is the $n$-dimensional zero vector, namely
		\begin{equation}
			\label{cond2}
			\ker(A) \cap \{ \mathbf{x} \in \mathbb{R}^n \mid W\mathbf{x} \in \operatorname{Im}(A^T) \} = \{ \mathbf{0} \}.
		\end{equation}
	\end{itemize}
\end{prop}

\noindent
{\bf Proof}.  
It is well known that a square matrix is invertible if and only if the associated homogeneous system has the zero vector as its unique solution. We thus analyze the homogeneous system $L\mathbf{v} = \mathbf{0}$, where the latter is the zero vector of $\mathbb{R}^{m+n}$, by decomposing $\mathbf{v} = \begin{pmatrix} \mathbf{y} \\ \mathbf{x} \end{pmatrix}$, with $\mathbf{y} \in \mathbb{R}^m$ and $\mathbf{x} \in \mathbb{R}^n$, to match the block structure (\ref{L}). This yields
\begin{equation}
	\label{lvdecomp}
	L\mathbf{v} = \mathbf{0} \quad \Leftrightarrow \quad 
	\left\{ 
	\begin{array}{l}
		A\mathbf{x} = \mathbf{0}, \\
		\\
		A^T\mathbf{y} + W\mathbf{x} = \mathbf{0},
	\end{array}
	\right.
\end{equation}
where the two zero vectors belong to $\mathbb{R}^m$ and $\mathbb{R}^n$, respectively.

\noindent
We now prove the sufficiency of the conditions. Assuming that conditions (\ref{cond1}) and (\ref{cond2}) hold, we show that $\mathbf{v} = \mathbf{0}$. From the two equations of system (\ref{lvdecomp}), we deduce
$$
\left\{ 
\begin{array}{l}
	A\mathbf{x} = \mathbf{0} \implies \mathbf{x} \in \ker(A), \\
	\\
	W\mathbf{x} = -A^T \mathbf{y} \implies W\mathbf{x} \in \operatorname{Im}(A^T).
\end{array}
\right.
$$
Combining these two results, it follows that $\mathbf{x}$ is a vector in the kernel of $A$ whose image $W\mathbf{x}$ belongs to the row space of $A$. By condition (\ref{cond2}), the only vector in $\mathbb{R}^n$ with this property is the zero vector, i.e., $\mathbf{x} = \mathbf{0}$. Substituting this into the second equation of system (\ref{lvdecomp}) yields $A^T\mathbf{y} = \mathbf{0}$. By virtue of condition (\ref{cond1}), $A^T$ has full column rank $m$ and is therefore injective, as $\dim(\ker(A^T)) = m - \operatorname{rank}(A^T) = 0$ by the rank-nullity Theorem. Consequently, we obtain $\mathbf{y} = \mathbf{0}$ as well. Thus, the unique solution to $L\mathbf{v} = \mathbf{0}$ is the zero vector $\mathbf{v} = \mathbf{0}$ in $\mathbb{R}^{m+n}$.

\noindent
We now address the necessity of the conditions: assuming $L$ is invertible, we show that both (\ref{cond1}) and (\ref{cond2}) necessarily hold. We proceed by contraposition: if either condition fails, then $L$ has a non-trivial kernel. If (\ref{cond1}) fails, i.e., $\operatorname{rank}(A)<m$, then there exists an $m$-dimensional non-zero vector $\mathbf{y}$ such that $A^T\mathbf{y}=\mathbf{0}$. Choosing $\mathbf{x}=\mathbf{0}$, the non-zero vector $\mathbf{v} = \begin{pmatrix} \mathbf{y} \\ \mathbf{x} \end{pmatrix}$ satisfies $L\mathbf{v}=\mathbf{0}$; hence, $L$ is singular.

\noindent
If, on the other hand, condition (\ref{cond2}) fails, there exists a non-zero vector $\mathbf{x} \neq \mathbf{0}$ belonging to the kernel of $A$ such that $W\mathbf{x}\in \operatorname{Im}(A^T)$. By definition, there exists $\mathbf{y}\in \mathbb{R}^m$ such that $W\mathbf{x}=-A^T\mathbf{y}$, or equivalently $A^T\mathbf{y}+W\mathbf{x}=\mathbf{0}$. The vector $\mathbf{v} = \begin{pmatrix} \mathbf{y} \\ \mathbf{x} \end{pmatrix}$ is non-zero and satisfies $L\mathbf{v}=\mathbf{0}$; thus, $L$ is singular.
$\quad \square$

\subsection{Geometric reading of the invertibility conditions}

\noindent
The first condition (\ref{cond1}) is immediate: requiring $A$ to have full row rank is equivalent to demanding that the $m$ constraint equations be linearly independent, ruling out redundant constraints or kinematic singularities in the given configuration.
If the rank of $A$ is strictly less than $m$, there exist at least two linearly dependent rows in $A$. The presence of the $m \times m$ zero block adjacent to $A$ (see (\ref{L})) ensures that this linear dependence transfers directly to the corresponding rows of the augmented matrix $L$, which is therefore singular.
From the standpoint of the information conveyed by the constraint conditions (\ref{vinc}), we see that matrix $A$ is precisely the one required to have full rank in order for the constraints to be independent: full rank thus ensures that there are no duplicate or redundant conditions in (\ref{vinc}).

\noindent
The second condition (\ref{cond2}) has a simple and clear geometric interpretation: there must be no non-zero vectors belonging to the kernel of $A$ that are mapped by the operator $W$ into vectors belonging to the orthogonal complement $(\ker(A))^\perp$. To establish this, relation (\ref{imort}) is taken into account, which here assumes the form $\operatorname{Im}(A^T)=(\ker(A))^\perp$. In other words, a vector in $\ker(A)$ cannot be ``rotated'' by $W$ into a vector orthogonal to the entire kernel $\ker(A)$.
Since the condition is both necessary and sufficient - assuming, of course, that (\ref{cond1}) holds - it represents a geometric characterization of the problem, serving as the minimal indispensable requirement to ensure that the overall linear system (\ref{eqmotolambda}) has a trivial kernel. The following example illustrates the situation in a simple case, enabling a clear geometric visualization.

\begin{exe}
Consider the case $n=3$ and $m=1$, defined by a single constraint matrix $A = \begin{pmatrix} a_1 & a_2 & a_3 \end{pmatrix} \neq {\bf 0}$ (hence $\operatorname{rank}(A) = 1$) and a diagonal matrix $W = \operatorname{diag}(\alpha_1, \alpha_2, \alpha_3)$, so that the matrix (\ref{L}) is 
$$
\left(
\begin{array}{cccc}
	0 & a_1 & a_2 & a_3 \\
	a_1 & \alpha_1 & 0 & 0 \\
	a_2 & 0 & \alpha_2 & 0 \\
	a_3 & 0 & 0 & \alpha_3
	\end{array}
\right)
$$
In this setting, $\ker(A)$ is the two-dimensional plane $a_1 x_1 + a_2 x_2 + a_3 x_3 = 0$, whereas $\operatorname{Im}(A^T) = (\ker(A))^\perp$ is the one-dimensional subspace spanned by $(a_1, a_2, a_3)^T$; moreover, the set of vectors $W{\bf x}$ as ${\bf x}$ varies in $\mathbb{R}^3$ is the subspace of vectors $(\alpha_1 x_1, \alpha_2 x_2, \alpha_3 x_3)^T$, whose dimension is equal to the number of non-zero $\alpha_i$.
The condition ${\bf x} \in \ker(A) \cap \{ {\bf x}\,\mid \,W{\bf x}\in \operatorname{Im}(A^T)\}$ requires finding a vector ${\bf x} = (x_1, x_2, x_3)^T$ and a scalar multiplier $\lambda \in {\Bbb R}$ satisfying the system:
	\begin{equation}
		\begin{cases}
			a_1 x_1 + a_2 x_2 + a_3 x_3 = 0, \\
			\alpha_1 x_1 = -\lambda a_1, \\
			\alpha_2 x_2 = -\lambda a_2, \\
			\alpha_3 x_3 = -\lambda a_3.
		\end{cases}
		\label{ex1}
	\end{equation}
We distinguish the following cases:
\begin{itemize}
	\item \emph{Definite matrix $W$ ($\alpha_i > 0$ or $\alpha_i < 0$ for all $i=1,2,3$):}
	Expressing $x_i = -\lambda \frac{a_i}{\alpha_i}$ and substituting these expressions into the first equation of (\ref{ex1}) yields:
	$$
	-\lambda \left( \frac{a_1^2}{\alpha_1} + \frac{a_2^2}{\alpha_2} + \frac{a_3^2}{\alpha_3} \right) = 0.
	$$
	The term in parentheses is strictly positive or strictly negative, forcing $\lambda = 0$ and consequently ${\bf x} = {\bf 0}$. Thus, the intersection is trivial and the full matrix $L$ is invertible, as confirmed by the straightforward determinant calculation:
	\begin{equation}
		\label{detexe}
		\det(L) = -a_1^2 \alpha_2 \alpha_3 - a_2^2 \alpha_1 \alpha_3 - a_3^2 \alpha_1 \alpha_2 = -\alpha_1 \alpha_2 \alpha_3 \left( \frac{a_1^2}{\alpha_1} + \frac{a_2^2}{\alpha_2} + \frac{a_3^2}{\alpha_3} \right).
	\end{equation}
	The geometric interpretation introduced earlier leads to the same conclusion: if $W$ is definite, no non-zero vector ${\bf x}$ can be orthogonal to $W{\bf x}$, since ${\bf x}^T W {\bf x} \neq 0$ for every ${\bf x} \neq {\bf 0}$. In particular, if ${\bf x}$ lies in the plane $\ker(A)$, the mapping $W$ cannot align it with the vector orthogonal to the plane, i.e., map it into the orthogonal complement $(\ker(A))^\perp$.
	
	\item \emph{Indefinite matrix $W$ with all $\alpha_i \neq 0$:} A non-trivial solution ${\bf x} \neq {\bf 0}$ exists if and only if the coefficients satisfy $\sum_{i=1}^3 \frac{a_i^2}{\alpha_i} = 0$. In this scenario, choosing $\lambda = -1$ yields the non-zero vector ${\bf x} = \left(\frac{a_1}{\alpha_1}, \frac{a_2}{\alpha_2}, \frac{a_3}{\alpha_3}\right)^T$, which satisfies ${\bf x} \in \ker(A)$ and $W{\bf x}=(a_1, a_2, a_3)^T \in (\ker(A))^\perp$, thus belonging to the intersection in (\ref{cond2}). By the Proposition, $L$ cannot be invertible, which is confirmed by the vanishing of the determinant (\ref{detexe}). \\
	However, this singularity condition does not necessarily hold for all indefinite matrices; indeed, indefinite matrices $W$ can still yield an invertible matrix $L$, as shown by the simple example $a_1=1, a_2=0, a_3=0, \alpha_1=\alpha_2=1, \alpha_3=-1$, for which $\det(L)=1$.
	
	\item \emph{Singular matrix $W$:} In this case, at least one of the $\alpha_i$ is zero. From the first expression in (\ref{detexe}), it is clear that the determinant may be non-zero or vanish depending on the parameters involved. For instance, taking $a_2=a_3=0$ with any $a_1 \neq 0$, and $\alpha_1=0, \alpha_2=\alpha_3=1$, the determinant is non-zero and $L$ is invertible. Conversely, taking $a_1=0$ with any $a_2^2+a_3^2 \neq 0$, and $\alpha_1=0$ with arbitrary $\alpha_2, \alpha_3$, we find $\det(L)=0$.
\end{itemize}

\noindent
The geometric interpretation in terms of $\ker(A)$ is straightforward in each of these cases. For instance, in the last scenario with $a_1=0$ and $\alpha_1=0$, the plane defining $\ker(A)$ is spanned by the vectors $(1,0,0)^T$ and $(0,-a_3, a_2)^T$ (a plane containing the $x_1$-axis) and is orthogonal to $(0,a_2, a_3)^T$, which spans $(\ker(A))^\perp$. On the other hand, the $x_1$-axis is the kernel of $W$; thus, any vector ${\bf x}=(\lambda, 0, 0)^T$ with $\lambda \in {\Bbb R}$ satisfies ${\bf x}\in \ker(A)$ and simultaneously $W{\bf x}={\bf 0} \in (\ker(A))^\perp$. Consequently, the intersection referred to in (\ref{cond2}) contains non-zero vectors, consistent with the singularity of the matrix $L$.
\end{exe}

\subsection{A physical interpretation relating to the motion of constrained systems}

\noindent
In addition to the geometric reading of the two conditions presented in Proposition 1, we provide here a physical interpretation based on the role played by matrices $A$ and $W$ in the context of constrained system dynamics.

\noindent
The two matrices $A$ and $W$ play distinct yet complementary roles:
while matrix $A$ encodes the kinematic restrictions imposed on the system, matrix $W$ characterizes its intrinsic inertial dynamics (in case (\ref{m}), defined by the Hessian of the Lagrangian function ${\cal L}$ with respect to velocities). Still in reference to case (\ref{m}) concerning mechanical system motion, the vector ${\bf x}$ represents the generalized acceleration vector $\ddot{\bf q}$. Consequently, the term $W{\bf x}$ plays the role of the generalized inertial forces associated with acceleration, representing the pure dynamic response of the unconstrained system.
While condition (\ref{cond1}) asserts the independence of the imposed kinematic constraints, condition (\ref{cond2}) governs how the system's inertia interacts with the motion allowed by the constraints.
Recalling (\ref{spostvirt}), we see that the kernel of $A$ defines the space of admissible motions, representing the directions along which the system can move freely without violating the constraints. Conversely, in light of the rule (\ref{imort}), the set $\operatorname{Im}(A^T)=(\ker(A))^\perp$ represents the space of constraint reaction forces, which act orthogonally to virtual displacements in agreement with the D'Alembert-Lagrange principle for ideal constraints.
The condition $W{\bf x}\in \operatorname{Im}(A^T)$ implies that this vector is orthogonal to the entire space of admissible motions.

\noindent
Let us examine what would happen if the intersection appearing in (\ref{cond2}) were non-trivial.
If there existed a non-zero vector ${\bf x}$ in this intersection, there would also exist a multiplier vector ${\bm \lambda}\in {\Bbb R}^m$ such that $W{\bf x}=-A^T {\bm \lambda}$ and, simultaneously, $A{\bf x}={\bf 0}$. The vector ${\bf x}$ would represent a kinematically admissible motion whose inertial forces are ``invisible'' to the effective degrees of freedom. The inertial effect of this motion would be absorbed entirely and exclusively by the constraint reaction forces ($\operatorname{Im}(A^T)$). Consequently, it would be possible to superimpose the vector ${\bf x}$ onto any motion solution without altering the applied external forces, leaving the system unable to ``perceive'' or uniquely determine this state of motion.
In other words, the force $W{\bf x}$ generated by the motion ${\bf x}$ points entirely in the direction of the constraint reactions $-A^T{\bm \lambda}$. As a result, this force has no usable component along the degrees of freedom spanned by $\ker(A)$; it is a force completely ``absorbed'' and suppressed by the constraints.

\noindent
Therefore, condition (\ref{cond2}) guarantees that no kinematically admissible motions exist that are ``dynamically invisible'' or completely decoupled from the equations of motion.
No ``ghost'' motions exist, and the dynamics are uniquely determined.
Otherwise, the resulting inertial force $W{\bf x} = -A^T {\bm \lambda} \in \operatorname{Im}(A^T)$ would be fully absorbed by the constraint reactions, discharging the system's inertia entirely onto the constraints and leaving the motion dynamically undetectable along the free directions.

\begin{rem}
	In the standard case of classical Lagrangians, we know that the matrix $W$ defined via (\ref{m}) is positive definite---a condition that, as we shall see, implies (\ref{cond2}) and is therefore sufficient for the invertibility of $L$. As already mentioned in the Introduction, however, there is no shortage of physical examples in which $W$ lacks this property (such as systems with massless physical components, or in the context of relativity); it is precisely in these settings that our considerations regarding the physical interpretation of the condition apply.
	More generally, these considerations can be extended to the structure (\ref{Lintr}), which encompasses various modeling scenarios not necessarily tied to mechanical motion.
	The block $W$ represents the energy or dynamics of the system and must be ``energetically'' active at least along the free directions permitted by the constraints $A$.
\end{rem}

\section{Further remarks, examples, and comparison with literature}

\noindent
In Example 1 of Paragraph 2.3 we saw how the properties of matrix $W$ can determine whether condition (\ref{cond2}) holds or not. We now wish to explore this aspect more systematically in order to clarify the link between the algebraic properties of $W$ and the invertibility of $L$. At the same time, by placing condition (\ref{cond2}) at the center of our analysis, we aim to understand how various other invertibility conditions found in the literature relate to it.

\noindent
The goal of this section is thus twofold:
\begin{itemize}
	\item to present a series of remarks, comments, and examples that help to better contextualize condition~(\ref{cond2}), which characterizes the non-singularity of matrix $L$;
	\item to compare this condition with those standardly reported in the literature.
\end{itemize}

\subsection{An investigation into condition~(\ref{cond2})}

\subsubsection{A necessary condition on the null spaces of $A$ and $W$}

\noindent
We begin by observing that (\ref{cond2}) implies that $W$ acts injectively on $\ker(A)$. Indeed, the following result holds.
\begin{cor}
	Condition (\ref{cond2}) implies
	\begin{equation}
		\label{wa}
		\ker(W) \cap \ker(A) = \{\mathbf{0}\}.
	\end{equation}
\end{cor}

\noindent
{\bf Proof}. 
Let $\mathbf{x} \in \ker(W) \cap \ker(A)$. Since $\mathbf{x} \in \ker(W)$, we have $W\mathbf{x} = \mathbf{0} \in \operatorname{Im}(A^T)$. Combining this with $\mathbf{x} \in \ker(A)$, it follows that $\mathbf{x}$ belongs to the set on the left-hand side of (\ref{cond2}). By virtue of (\ref{cond2}), we conclude that $\mathbf{x} = \mathbf{0}$.
$\quad \square$

\noindent
Since the intersection of the kernel spaces is trivial, the only vector in $\ker(A)$ mapped to zero by $W$ is the zero vector itself. Consequently, the restriction of $W$ to $\ker(A)$ has a trivial kernel, meaning it is an injective mapping.

\noindent
Furthermore, condition (\ref{cond2}) prevents a non-zero vector $\mathbf{x} \in \ker(A)$ from being mapped by $W$ into a non-zero vector of $\operatorname{Im}(A^T)$.

\begin{rem}
	Condition (\ref{wa}) is necessary for the invertibility of $L$, but not sufficient (and thus it is not equivalent to (\ref{cond2})). This is precisely because it cannot prevent the "absorption" of a vector $\mathbf{x} \in \ker(A)$ into $\operatorname{Im}(A^T)$ after being transformed by $W$.
\end{rem}
	
\noindent
A simple counterexample illustrates this point.

\begin{exe}
For $m=1$ and $n=2$, let $A = \begin{pmatrix} 1 & 0 \end{pmatrix}$ and $W = \begin{pmatrix} 0 & 1 \\ 0 & 0 \end{pmatrix}$. The resulting matrix $L$ defined in (\ref{L}),
	\[
	L = \begin{pmatrix}
		0 & 1 & 0 \\
		1 & 0 & 1 \\
		0 & 0 & 0
	\end{pmatrix},
	\]
is clearly singular. Nevertheless, condition (\ref{wa}) holds: indeed, $\ker(W) = \operatorname{span}\left\{ \begin{pmatrix} 1 \\ 0 \end{pmatrix} \right\}$ and $\ker(A) = \operatorname{span}\left\{ \begin{pmatrix} 0 \\ 1 \end{pmatrix} \right\}$, so that $\ker(W) \cap \ker(A) = \{\mathbf{0}\}$. 
On the other hand, we can verify that condition (\ref{cond2}) does not hold (which avoids any contradiction with Proposition 2): choosing $\bar{\mathbf{x}} = \begin{pmatrix} 0 \\ 1 \end{pmatrix} \in \ker(A)$, we compute $W\bar{\mathbf{x}} = \begin{pmatrix} 1 \\ 0 \end{pmatrix}$. Since $\operatorname{Im}(A^T) = \operatorname{span}\left\{ \begin{pmatrix} 1 \\ 0 \end{pmatrix} \right\}$, it follows that $W\bar{\mathbf{x}} \in \operatorname{Im}(A^T)$. Thus, the set on the left-hand side of (\ref{cond2}) contains the non-zero vector $\bar{\mathbf{x}}$. We also note that in the non-singular case $\det W \neq 0$, condition (\ref{wa}) holds trivially, since $\ker(W)$ reduces to the zero vector $\{\mathbf{0}\}$ alone. However, there exist cases where $W$ is non-singular but the full matrix $L$ is singular (as we will illustrate shortly), confirming that condition (\ref{wa}) cannot be sufficient for the invertibility of $L$.
\end{exe}

\begin{rem}
If there exists a non-zero vector ${\bf x}$ belonging to the intersection $\ker(W) \cap \ker(A)$, then condition (\ref{cond2}) automatically fails, since (\ref{wa}) would be necessary for the latter to hold. On the other hand, the failure of (\ref{cond2}) is clear, given that $W{\bf x} = {\bf 0} \in (\ker(A))^\perp$,thus the non-zero vector ${\bf x}$ would belong to the kernel of $A$ and be mapped by $W$ into $\operatorname{Im}(A^T)$.
\end{rem}

\subsubsection{The role of matrix $W$}

\noindent
We emphasize that the criterion formulated in Proposition 2 does not require $W$ to be an invertible matrix: whether $W$ is singular or non-singular is independent of the singularity of the full matrix $L$, as illustrated by the following two examples.

\begin{exe}
	For $n=2$ and $m=1$, choosing $A=\begin{pmatrix} 1 & 0 \end{pmatrix}$ and $W=\begin{pmatrix} 0 & 0 \\ 0 & 1 \end{pmatrix}$, $W$ is singular, yet $L$ is non-singular with $\det L = -1$. On the other hand, the two conditions (\ref{cond1}) and (\ref{cond2}) are satisfied (as expected, since $L$ is invertible), as $A$ has full rank, $\operatorname{Im}(A^T)=\operatorname{span}\left\{ \begin{pmatrix} 1 \\ 0 \end{pmatrix} \right\}$, and $W\mathbf{x} = \begin{pmatrix} 0 \\ c \end{pmatrix} \notin \operatorname{Im}(A^T)$ for every non-zero $\mathbf{x} = \begin{pmatrix} 0 \\ c \end{pmatrix} \in \ker(A)$.
	So to speak, the singularity of $W$ lies outside the kernel of $A$ in a way that does not introduce a kernel for the full block matrix $L$.
\end{exe}

\begin{exe}
	As a second example, we fix $m=2$ and $n=3$ and choose $A=\begin{pmatrix} 1 & 0 & 0 \\ 0 & 0 & 1 \end{pmatrix}$ and $W=\begin{pmatrix} 1 & 1 & 0 \\ 1 & 1 & 1 \\ 0 & 1 & 0 \end{pmatrix}$. Clearly, $A$ has full rank equal to $2$ and $\det W = -1$, meaning $W$ is non-singular. However, the full matrix $L$ is singular, as its second and fifth rows are identical, which can be easily verified. Indeed, in this case condition (\ref{cond2}) fails, since
	$\ker(A) = \operatorname{span}\left\{ \begin{pmatrix} 0 \\ 0 \\ 1 \end{pmatrix} \right\}$, 
	$\operatorname{Im}(A^T) = \left\{ \begin{pmatrix} \alpha \\ 0 \\ \beta \end{pmatrix} \in \mathbb{R}^3 \middle\vert\ \alpha, \beta \in \mathbb{R} \right\}$, and
	$\{ \mathbf{x} \in \mathbb{R}^3 \mid W\mathbf{x} \in \operatorname{Im}(A^T) \} = \left\{ \begin{pmatrix} x_1 \\ 0 \\ x_3 \end{pmatrix} \in \mathbb{R}^3 \middle\vert\ x_1, x_3 \in \mathbb{R} \right\}$.
	Therefore,
	$\ker(A) \cap \{ \mathbf{x} \in \mathbb{R}^3 \mid W\mathbf{x} \in \operatorname{Im}(A^T) \} = \ker(A) = \left\{ \begin{pmatrix} 0 \\ 0 \\ c \end{pmatrix} \middle\vert\ c \in \mathbb{R} \right\} \neq \{ \mathbf{0} \}$.
\end{exe}

\begin{rem}
The potential singularity of matrix $W$ explains why condition (\ref{cond2}) cannot be formulated on the image spaces by applying $W$ to the spaces appearing in (\ref{cond2}) to obtain
\begin{equation}
	\label{condimm}
	W(\ker(A)) \cap \operatorname{Im}(A^T) = \{ \mathbf{0} \}.
\end{equation}
If $W$ is non-singular, (\ref{cond2}) and (\ref{condimm}) are equivalent. However, if $W$ is singular, $W\mathbf{x} = \mathbf{0}$ does not imply $\mathbf{x} = \mathbf{0}$. The transformation $W$ maps the entire kernel $\ker(W)$ to the zero vector $\mathbf{0}$, thereby hiding, so to speak, the potential presence of non-zero vectors $\mathbf{x} \in \ker(A) \cap \ker(W)$. Condition (\ref{condimm}) is therefore necessary, but in general not sufficient, for the invertibility of $L$.
\end{rem}

\noindent
For completeness, although not strictly necessary for the scope of our discussion, we can establish a scheme of implications and relationships among the three conditions that emerged: (\ref{cond2}) (which, together with (\ref{cond1}), is necessary and sufficient for the invertibility of $L$), and (\ref{wa}) and (\ref{condimm}) (which are only necessary):
	$$
	\begin{array}{ll}
		(\ref{cond2})\,\Rightarrow\,(\ref{wa}), & (\ref{cond2})\,\Rightarrow\,(\ref{condimm}),\\[1ex]
		(\ref{wa})\,\not \Rightarrow \,(\ref{condimm}), &(\ref{condimm})\,\not \Rightarrow \,(\ref{wa}), \\[1ex]
		(\ref{wa})\,\wedge \,(\ref{condimm})\,\Leftrightarrow\,(\ref{cond2}).
	\end{array}
	$$
For the last statement, we can reason as follows: let $\mathbf{x} \in \ker(A)$ be such that $W\mathbf{x} \in \operatorname{Im}(A^T)$. Then $W\mathbf{x} \in W(\ker(A)) \cap \operatorname{Im}(A^T) = \{\mathbf{0}\}$ (using (\ref{condimm})), which means $W\mathbf{x} = \mathbf{0}$. Consequently, $\mathbf{x} \in \ker(A) \cap \ker(W) = \{\mathbf{0}\}$ (using (\ref{wa})), from which $\mathbf{x} = \mathbf{0}$.

\subsubsection{A comparison with the standard positivity condition on $W$}

\noindent
We now clarify how the assumption that the matrix $W$ has a definite sign (positive or negative) on the entire space $\mathbb{R}^n$ or on a subspace thereof affects the invertibility of $L$, and we compare this with previous results. We first establish the following result.

\begin{prop}
	If $W$ is sign-definite on the subspace $\ker(A)$, that is,
	\begin{equation}
		\label{fq}
\mathbf{x}^T W \mathbf{x} > 0\;\; or \;\;\mathbf{x}^T W \mathbf{x} < 0 \quad \forall \mathbf{x} \in \ker(A) \setminus \{\mathbf{0}\},
	\end{equation}
	then condition~\eqref{cond2} holds.
\end{prop}

\noindent
{\bf Proof}.
Recall property (\ref{imort}), which yields $\operatorname{Im}(A^T) = (\ker(A))^\perp$; that is, a vector $\mathbf{y}$ belongs to the image of $A^T$ if and only if $\mathbf{y}$ is orthogonal to every vector in the kernel of $A$:
	\begin{equation}
		\label{imperp}
		\mathbf{y} \in \operatorname{Im}(A^T) \;\iff\; \mathbf{y}^T \mathbf{z} = 0 \quad \forall \mathbf{z} \in \ker(A).
	\end{equation}
	Let $\mathbf{x} \in \ker(A)$ be such that $W\mathbf{x} \in \operatorname{Im}(A^T)$. By (\ref{imperp}), $W\mathbf{x} \in (\ker(A))^\perp$, which implies that $W\mathbf{x}$ is orthogonal to $\mathbf{x}$ as well:
	\[
	\mathbf{x}^T W \mathbf{x} = 0.
	\]
	Since $\mathbf{x} \in \ker(A)$, assumption (\ref{fq}) applies and implies that $\mathbf{x} = \mathbf{0}$. Therefore, condition (\ref{cond2}) holds (the intersection contains only the zero vector).
$\quad\square$

\noindent
We emphasize that it is neither required for $W$ to be symmetric, nor for $\det(W) \neq 0$, nor for $W$ to have a definite sign on the whole space $\mathbb{R}^n$. The matrix $W$ may even be indefinite overall, provided that its restriction to the kernel of $A$ is definite (positive or negative), while its indefinite component acts on the remaining subspace.

\begin{exe}
	For $n=3$ and $m=1$, let $A = (0 \; 0 \; 1)$ and $W = \begin{pmatrix} 2 & 0 & 0 \\ 0 & 3 & 0 \\ 0 & 0 & -5 \end{pmatrix}$, which is indefinite on $\mathbb{R}^3$.
	However, we have
	\[
	\ker(A) = \left\{ \mathbf{x} = \begin{pmatrix} a \\ b \\ 0 \end{pmatrix} \;\middle|\; a, b \in \mathbb{R} \right\},
	\]
	and the quadratic form restricted to the kernel satisfies $\mathbf{x}^T W \mathbf{x} \vert_{\ker(A)} = 2a^2 + 3b^2 > 0$ for all $\mathbf{x} \neq \mathbf{0}$. Condition (\ref{cond2}) indeed holds, since
	\[
	W\mathbf{x} \vert_{\ker(A)} = \begin{pmatrix} 2a \\ 3b \\ 0 \end{pmatrix} \quad \text{and} \quad \operatorname{Im}(A^T) = \operatorname{span} \left\{ \begin{pmatrix} 0 \\ 0 \\ 1 \end{pmatrix} \right\},
	\]
	which intersect solely at $\mathbf{x} = \mathbf{0}$.
\end{exe}

\noindent
We also underline that condition (\ref{fq}) is sufficient but not necessary, meaning that the converse implication $(\ref{cond2}) \implies (\ref{fq})$ does not hold in general, as illustrated by the following example.

\begin{exe}
	Let $A = (0 \; 0 \; 1)$ and $W = \begin{pmatrix} 1 & 0 & 0 \\ 0 & -1 & 0 \\ 0 & 0 & 0 \end{pmatrix}$. The subspaces $\ker(A)$ and $\operatorname{Im}(A^T)$ are the same as in the previous example, while
	\[
	W\mathbf{x} \vert_{\ker(A)} = \begin{pmatrix} a \\ -b \\ 0 \end{pmatrix},
	\]
	which belongs to $\operatorname{Im}(A^T)$ if and only if $a = b = 0$; thus, condition (\ref{cond2}) holds. However, evaluating the quadratic form on the kernel yields $\mathbf{x}^T W \mathbf{x} \vert_{\ker(A)} = a^2 - b^2$, which vanishes for non-zero vectors as well, such as $(1 \; 1 \; 0)^T \in \ker(A)$.
\end{exe}

\noindent
The following statement, which follows from Proposition 3, provides a direct comparison between the results achieved so far and those related to the standard assumption in the literature (also adopted in \cite{lemos}) of a positive definite matrix $W$:

\begin{cor}
	If the matrix $W$ is definite (positive or negative) on the entire space $\mathbb{R}^n$, then assumption (\ref{cond1}) alone ensures that the block matrix $L$ in (\ref{L}) is invertible.
\end{cor}

\begin{proof}
	Clearly, if $\mathbf{y}^T W \mathbf{y} \neq 0$ for all $\mathbf{y} \in \mathbb{R}^n \setminus \{\mathbf{0}\}$, the same property holds when restricted to the kernel $\ker(A)$, and thus (\ref{fq}) holds.
\end{proof}

\noindent
The preceding Corollary serves as the bridge between the sufficient-type results standard in the literature and those arising from our analysis, which address the necessary and sufficient characterization for the invertibility of $L$. The fact that a definite sign is not necessary is easily illustrated by a straightforward example.

\begin{exe}
Consider for $m=1$ and $n=2$ the matrix
$$
L=\left( \begin{array}{ccc} 0 & 1 & 0 \\ 1 & \alpha & 0 \\ 0 & 0 & 1  \end{array} \right)
$$
where $A=(1 , 0)$ and $W=\left( \begin{array}{cc}\alpha & 0 \\ 0 & 1 \end{array} \right)$. Since $\det L =-1$, the matrix is always invertible for any value of $\alpha \in \mathbb{R}$; however, matrix $M$ is positive definite only for $\alpha>0$ (for $\alpha=0$ it is singular, and for $\alpha<0$ it is indefinite).
\end{exe}
 
\subsubsection{The case of symmetric $W$}

The symmetry property of the matrix $W$ - which is certainly common, starting from the equations of motion~\eqref{m} where $M_{i,j}=M_{j,i}$ - allows one to refine the invertibility criterion, provided that $W$ is sign-semidefinite. Indeed, the following result holds.

\begin{lem}
	Let $W \in \mathbb{R}^{n \times n}$ be a symmetric and sign-semidefinite matrix (i.e., $W \ge 0$ or $W \le 0$). If
	\begin{equation*}
		\ker (W) \cap \ker (A) = \{\mathbf{0}\},
	\end{equation*}
	then
	\begin{equation*}
		\ker (A) \cap \{ \mathbf{x} \in \mathbb{R}^n \mid W\mathbf{x} \in \mathrm{im}(A^T) \} = \{\mathbf{0}\}
	\end{equation*}
that is (\ref{wa})$\,\Rightarrow \,$(\ref{cond2}).
\end{lem}

\noindent
{Proof}.
Let $\mathbf{x} \in \ker (A) \cap \{ \mathbf{x} \in \mathbb{R}^n \mid W\mathbf{x} \in \mathrm{im}(A^T) \}$. By definition, $\mathbf{x}$ satisfies:
\begin{itemize}
		\item[$(i)$] $A\mathbf{x} = \mathbf{0}$,
		\item[$(ii)$] $W\mathbf{x} = A^T \mathbf{y}$ for some vector $\mathbf{y} \in \mathbb{R}^m$.
	\end{itemize}
Multiplying condition (ii) from the left by $\mathbf{x}^T$ and substituting condition (i), we obtain:
	\begin{equation}
		\mathbf{x}^T W \mathbf{x} = \mathbf{x}^T A^T \mathbf{y} = (A\mathbf{x})^T \mathbf{y} = \mathbf{0}^T \mathbf{y} = 0.
	\end{equation}
Since $W$ is symmetric and sign-semidefinite (assuming $W \ge 0$ without loss of generality; if $W \le 0$, one considers $-W \ge 0$), the vanishing of the quadratic form $\mathbf{x}^T W \mathbf{x} = 0$ implies $W\mathbf{x} = \mathbf{0}$, meaning that $\mathbf{x} \in \ker (W)$.
Consequently, $\mathbf{x} \in \ker (W) \cap \ker (A)$. By the hypothesis $\ker (W) \cap \ker (A) = \{\mathbf{0}\}$, it follows that $\mathbf{x} = \mathbf{0}$, which proves (\ref{cond2}).$\quad\square$

\noindent
Note that condition (\ref{cond2}) always implies (\ref{wa}), regardless of any structural properties of $W$, as previously observed at the end of Paragraph~3.1.2. Therefore, the following statement holds:
\begin{prop}
Under the assumption that $W$ is symmetric and sign-semidefinite, condition (\ref{cond2}) and condition (\ref{wa}) are fully equivalent:
\begin{equation}
	\label{cond2wa}
	\begin{array}{l}
		\ker (A) \cap \{ \mathbf{x} \in \mathbb{R}^n \mid W\mathbf{x} \in \mathrm{im}(A^T) \} = \{\mathbf{0}\}
\;\Longleftrightarrow\;	\ker (W) \cap \ker (A) = \{\mathbf{0}\}
\\
\\
\textrm{if}\;W\;\textrm{is symmetric and sign-semidefinite.}
\end{array}.
	\end{equation}
\end{prop}

\begin{rem}
To appreciate the necessity of the symmetry assumption on $W$, consider the unique decomposition of $W$ into its symmetric and skew-symmetric parts:
	\begin{equation*}
		W = W_S + W_A, \quad \text{where } W_S = \frac{W + W^T}{2} \text{ and } W_A = \frac{W - W^T}{2}.
	\end{equation*}
Since $\mathbf{x}^T W_A \mathbf{x} = 0$ for any vector $\mathbf{x}$, the quadratic form depends exclusively on the symmetric part, i.e., $\mathbf{x}^T W \mathbf{x} = \mathbf{x}^T W_S \mathbf{x}$. 
In the proof of the Lemma, the condition $\mathbf{x}^T W \mathbf{x} = 0$ yields $\mathbf{x}^T W_S \mathbf{x} = 0$, which implies $W_S \mathbf{x} = \mathbf{0}$ (under the sign-semidefiniteness of $W_S$). However, if $W$ is not symmetric (so $W_A \neq \mathbf{0}$), this does not imply $W_A \mathbf{x} = \mathbf{0}$. Consequently, one obtains
	\begin{equation*}
		W \mathbf{x} = W_S \mathbf{x} + W_A \mathbf{x} = W_A \mathbf{x},
	\end{equation*}
which can be non-zero. If this non-zero vector $W_A \mathbf{x}$ happens to belong to $\mathrm{im}(A^T)$, then $\mathbf{x}$ belongs to the subspace in condition (\ref{cond2}) without belonging to $\ker (W)$, causing the implication $\eqref{wa} \implies \eqref{cond2}$ to fail.
\end{rem}

\noindent
For a symmetric matrix $W$ it is equally interesting to note that condition (\ref{cond2}) implies (\ref{fq}) (and thus they are equivalent), provided again that $W$ is sign-semidefinite. Indeed, we have the following:

\begin{lem}
	Let $W \in \mathbb{R}^{n \times n}$ be symmetric and sign-semidefinite. Then condition~\eqref{cond2} implies~\eqref{fq}.
\end{lem}

\begin{proof}
	Assume by contradiction that there exists $\mathbf{y} \in \ker (A) \setminus \{\mathbf{0}\}$ such that $\mathbf{y}^T W \mathbf{y} = 0$. By a standard property of symmetric sign-semidefinite matrices, $\mathbf{y}^T W \mathbf{y} = 0$ implies $W\mathbf{y} = \mathbf{0}$.
	Since $W\mathbf{y} = \mathbf{0} = A^T \mathbf{0} \in \mathrm{im}(A^T)$, it follows that $W\mathbf{y} \in \mathrm{im}(A^T)$ for a non-zero vector $\mathbf{y} \in \ker (A)$. This directly contradicts condition~\eqref{cond2}. We conclude that~\eqref{fq} must hold under the stated assumptions on $W$.
\end{proof}

\subsubsection{An equivalent condition for the invertibility of $L$}

\noindent
Assuming (\ref{cond1}), we construct a matrix $V$ whose columns consist of $n-m$ basis vectors for $\ker(A)$. We prove the following result.

\begin{prop}
	Let $V \in \mathbb{R}^{n \times (n-m)}$ be a matrix whose columns form a basis for $\ker(A)$. Condition~(\ref{cond2}) holds if and only if 
	\begin{equation}
		\label{condeq}
		\det(V^T W V) \neq 0.
	\end{equation}
\end{prop}

\noindent
{\bf Proof}. 
Recall that $\operatorname{Im}(A^T) = (\ker(A))^\perp$. Any vector $\mathbf{x} \in \ker(A)$ can be uniquely written as $\mathbf{x} = V \mathbf{c}$ for some $\mathbf{c} \in \mathbb{R}^{n-m}$, where $\mathbf{x} \neq \mathbf{0}$ if and only if $\mathbf{c} \neq \mathbf{0}$.
Condition (\ref{cond2}) fails to hold if and only if there exists $\mathbf{x} \in \ker(A) \setminus \{\mathbf{0}\}$ such that $W\mathbf{x} \in \operatorname{Im}(A^T)$, which is equivalent to $W\mathbf{x} \perp \ker(A)$. Expressed in terms of the basis matrix $V$, this orthogonality condition translates to $V^T (W\mathbf{x}) = \mathbf{0}$.
Substituting $\mathbf{x} = V\mathbf{c}$, condition (\ref{cond2}) fails if and only if there exists a non-zero vector $\mathbf{c} \in \mathbb{R}^{n-m}$ such that $V^T W V \mathbf{c} = \mathbf{0}$. The existence of such a vector $\mathbf{c}$ is equivalent to $\det(V^T W V) = 0$, which completes the proof.
$\quad \square$

\noindent
In practical terms, the matrix $V^T W V \in \mathbb{R}^{(n-m) \times (n-m)}$ represents the restriction of the bilinear form associated with $W$ to $\ker(A)$ with respect to the basis $V$. Consequently, condition (\ref{condeq}) states that this restriction is non-degenerate. This directly connects to condition~\eqref{fq}: while condition (\eqref{fq}) is a stronger requirement (implying that the quadratic form has no non-zero isotropic vectors in $\ker(A)$), it guarantees condition (\ref{condeq}) as a sufficient, albeit not necessary, condition.

It is further worth noting that the validity of this condition is independent of the choice of the basis matrix $V$ and that, as in the previous criteria, the invertibility of $W$ is not required. A trivial example of a singular matrix $W$ satisfying $\det(V^T W V) \neq 0$ can be constructed for $m=1$ and $n=2$ with $A = (1 \; 0)$ and $W = \begin{pmatrix} 0 & 0 \\ 0 & 1 \end{pmatrix}$, which is evidently singular: we find $V = \begin{pmatrix} 0 \\ 1 \end{pmatrix}$, yielding $\det(V^T W V) = 1$.

\subsubsection{Schur complement and Haynsworth rank formula}

\noindent
Let $M$ be a block matrix of size $(p+n) \times (q+n)$, partitioned as
\begin{equation}
	\label{mschur}
	M = \begin{pmatrix} M_{11} & M_{12} \\ M_{21} & M_{22} \end{pmatrix},
\end{equation}
where $M_{11} \in \mathbb{R}^{p \times q}$, $M_{12} \in \mathbb{R}^{p \times n}$, $M_{21} \in \mathbb{R}^{n \times q}$, and $M_{22} \in \mathbb{R}^{n \times n}$ is a non-singular matrix. The Schur complement of $M_{22}$ in $M$, denoted by $(M/M_{22})$, is a matrix of size $p \times q$ defined as (see \cite{zhang} for a detalied presentation)
\begin{equation}
	\label{schurcompl}
	(M/M_{22}) = M_{11} - M_{12} M_{22}^{-1} M_{21}.
\end{equation}
By virtue of the block LDU (Lower-Diagonal-Upper, see for instance \cite{bernstein, golub}) factorization of $M$, the following relations hold:
\begin{itemize}
	\item \textbf{Haynsworth Rank Formula:}
	\begin{equation}
		\label{hrf}
		\operatorname{rank}(M) = \operatorname{rank}(M_{22}) + \operatorname{rank}\left(M_{11} - M_{12} M_{22}^{-1} M_{21}\right).
	\end{equation}
\label{sdf}
	\item \textbf{Schur's Determinant Formula} (for the case $p = q$, where $M$ and $M_{11}$ are square matrices of size $(p+n) \times (p+n)$ and $p \times p$, respectively):
	\begin{equation}
		\det(M) = \det(M_{22}) \cdot \det\left(M_{11} - M_{12} M_{22}^{-1} M_{21}\right).
	\end{equation}
\end{itemize}
Consequently, assuming $p = q$ and $M_{22}$ non-singular, $M$ is non-singular if and only if its Schur complement (\ref{schurcompl}) is non-singular. For further details and a comprehensive overview of the Schur determinant formula and the Haynsworth rank formula we refer the reader to \cite{cottleschur, zhang}.

\noindent
Let us apply the result to the matrix $L$ defined in (\ref{L}).
Assuming that the block $W$ is non-singular, we can apply Schur's complement theory to the matrix $L$ defined in (\ref{L}) by setting $p=q=m$ and (with respect to (\ref{mschur})) $M_{11} = \mathbb{O}$, $M_{12} = A$, $M_{21} = A^T$, and $M_{22} = W$. The Schur complement of $W$ in $L$ is given by
\begin{equation}
	\label{lw}
	(L/W) = -A W^{-1} A^T.
\end{equation}
Consequently, by virtue of Haynsworth rank formula and Schur's determinant formula, we obtain from (\ref{hrf}) and (\ref{sdf}):
$$
\operatorname{rank}(L) = \operatorname{rank}(W) + \operatorname{rank}\left(A W^{-1} A^T\right),
\quad
\det(L) = (-1)^m \det(W) \cdot \det\left(A W^{-1} A^T\right).
$$
Therefore, provided that $W$ is non-singular, the full matrix $L$ is non-singular if and only if the matrix $A W^{-1} A^T \in \mathbb{R}^{m \times m}$ is non-singular.
The non-singularity of $A W^{-1} A^T$ must therefore be, in cases where $W$ is invertible, equivalent to the conditions established in Proposition 2; indeed, the following lemma holds:

\begin{lem}
	Let $L \in \mathbb{R}^{(m+n) \times (m+n)}$ be the matrix defined in (\ref{L}), with $W$ non-singular.
	Then, the Schur complement (\ref{lw}) is non-singular if and only if 
	the two conditions (\ref{cond1}) and (\ref{cond2}) are simultaneously satisfied.
\end{lem}

\noindent
{\bf Proof}.
Assume that $-A W^{-1} A^T$ is non-singular. By the properties of rank for matrix products, we have
$m = \operatorname{rank}(A W^{-1} A^T) \le \operatorname{rank}(A) \le m$, which implies $\operatorname{rank}(A) = m$. Furthermore, let $\mathbf{x} \in \ker(A) \cap \{ \mathbf{x} \in \mathbb{R}^n \mid W\mathbf{x} \in \operatorname{Im}(A^T) \}$. Since $W\mathbf{x} \in \operatorname{Im}(A^T)$, there exists a vector $\mathbf{y} \in \mathbb{R}^m$ such that $W\mathbf{x} = A^T \mathbf{y}$, which implies $\mathbf{x} = W^{-1} A^T \mathbf{y}$.
Applying $A$ to both sides yields $A\mathbf{x} = A W^{-1} A^T \mathbf{y}$.
Because $\mathbf{x} \in \ker(A)$, we have $A\mathbf{x} = \mathbf{0}$, so $A W^{-1} A^T \mathbf{y} = \mathbf{0}$. Since $A W^{-1} A^T$ is non-singular, it follows that $\mathbf{y} = \mathbf{0}$. Substituting $\mathbf{y} = \mathbf{0}$ back into the expression for $\mathbf{x}$, we obtain $\mathbf{x} = W^{-1} A^T \mathbf{0} = \mathbf{0}$. Thus, the intersection contains only the zero vector and (\ref{cond2}) holds.

\noindent
Conversely, assume that conditions (\ref{cond1}) and (\ref{cond2}) hold. We must show that $\ker(A W^{-1} A^T) = \{ \mathbf{0} \}$. Let $\mathbf{y} \in \mathbb{R}^m$ be such that $A W^{-1} A^T \mathbf{y} = \mathbf{0}$ and define $\mathbf{x} = W^{-1} A^T \mathbf{y} \in \mathbb{R}^n$.
Applying $A$ to $\mathbf{x}$, we have $A\mathbf{x} = A W^{-1} A^T \mathbf{y} = \mathbf{0}$, which implies $\mathbf{x} \in \ker(A)$. Applying $W$ to $\mathbf{x}$, we get $W\mathbf{x} = W(W^{-1} A^T \mathbf{y}) = A^T \mathbf{y}$, which implies $W\mathbf{x} \in \operatorname{Im}(A^T)$.
Therefore, $\mathbf{x} \in \ker(A) \cap \{ \mathbf{x} \in \mathbb{R}^n \mid W\mathbf{x} \in \operatorname{Im}(A^T) \}$. By Condition (\ref{cond2}), we must have $\mathbf{x} = \mathbf{0}$, which means $W^{-1} A^T \mathbf{y} = \mathbf{0}$, and hence $A^T \mathbf{y} = \mathbf{0}$.
Since $\operatorname{rank}(A) = m$, the matrix $A^T \in \mathbb{R}^{n \times m}$ has full column rank $m$, so its kernel is trivial. Thus, $A^T \mathbf{y} = \mathbf{0}$ implies $\mathbf{y} = \mathbf{0}$. 
We conclude that $\ker(A W^{-1} A^T) = \{ \mathbf{0} \}$, which is equivalent to the non-singularity of $A W^{-1} A^T$.
$\quad \square$

\begin{rem}
It is worth emphasizing that the characterization provided by conditions (\ref{cond1}) and (\ref{cond2}) is considerably more general. While the Schur complement approach yields information exclusively when the block $W$ is non-singular, the two-condition criterion remains applicable even when $W$ is singular (or even non-square, provided the dimensions match). Thus, it serves as a unified framework for characterizing the non-singularity of $L$ regardless of the invertibility of $W$. Relying solely on the Schur complement to characterize the general invertibility of $L$ would fail to provide a complete picture, as it overlooks the cases where $L$ is invertible due to the action of $A$, despite $W$ being singular or rank-deficient.
\end{rem}

\subsection{Relation to classical results}

\subsubsection{$W$ positive definite}

In classical systems with real, distributed or lumped positive masses, $W$ is positive definite over the entire space $\mathbb{R}^n$ ($x^T W x > 0, \, \forall x \neq \mathbf{0}$). In this scenario, $x^T W x > 0$ holds \emph{a fortiori} on $\ker (A)$, and the intersection condition is automatically satisfied for any constraint matrix $A$.

\noindent
Requiring $A$ to have full rank and $W$ to be positive definite is the most robust and straightforward approach to guarantee the invertibility of $L$. This is the assumption adopted, for instance, in \cite{lemos}, where the objective is to ensure that constraint forces can be uniquely determined (via Lagrange multipliers) strictly as functions of positions $\mathbf{q}$, velocities $\dot{\mathbf{q}}$, and time.

\noindent
The condition that $W$ is positive definite involves exclusively the structural and dynamical properties of the system, independently of the imposed system of constraint forces. That is, there is no interplay between the inertial and constraint components of the system that could relax the positive-definiteness assumption while still preserving the invertibility of the system.

\noindent
On the other hand, for classical mechanical systems with Lagrangian $\mathcal{L} = T - V$, the positive definiteness of the Hessian matrix~\eqref{m} is automatic provided that the masses are positive and the generalized coordinates are independent, given that $T = \frac{1}{2} \dot{\mathbf{q}}^T W \dot{\mathbf{q}}$.
More generally, in the literature - such as in the context of \cite{babuska, brezzi1974}, or LBB (Ladyzhenskaya-Babu{\v{s}}ka-Brezzi) conditions - one often assumes the sufficient condition that $W$ is strictly positive definite on the space of admissible motions, i.~e.~$x^T W x > 0, \, \forall x \in \ker (A) \setminus \{\mathbf{0}\}$ (corresponding to the coercivity condition on the kernel), which falls under the condition previously examined in (\ref{fq}).
In the context of the LBB theory for saddle-point problems, the standard stability requirements demand $W$ to be strictly coercive (positive definite) on $\ker(A)$. Condition~\eqref{fq} ($\mathbf{x}^T W \mathbf{x} \neq 0$ for all non-zero $\mathbf{x} \in \ker(A)$) relaxes this assumption by allowing $W$ to be definite (either positive or negative) on the kernel.

\noindent
It is important to emphasize that the algebraic condition derived here, $\ker (A) \cap \{x \mid Wx \in (\ker (A))^\perp\} = \{\mathbf{0}\}$, is more general and less restrictive: it does not require kinetic energy to be strictly positive along every direction of $\ker (A)$ (let alone on the entire space $\mathbb{R}^n$), but merely dictates that there exist no non-zero admissible motions whose generated inertial forces fall entirely orthogonal to the space of virtual motions.

\subsubsection{$W$ positive semidefinite}

\noindent
Singular mass matrices $W$ (possessing zero eigenvalues) typically arise in the presence of massless physical components, or when nonholonomic constraints are accommodated by introducing auxiliary variables devoid of inertia.
In such situations, the matrix $W$ is typically positive semidefinite. A criterion of type (\ref{fq}), namely $x^T W x > 0, \, \forall x \in \ker (A) \setminus \{\mathbf{0}\}$, no longer holds, as shown by the simple counterexample ($n=2, m=1$) with $A = \begin{pmatrix} 1 & 0 \end{pmatrix}$ and $W = \begin{pmatrix} 1 & 0 \\ 0 & 0 \end{pmatrix}$.

\noindent
In \cite{udwadia2006} the authors consider a dynamical system with a symmetric, positive semidefinite matrix $W$ and a constraint matrix $A$ not necessarily of full rank. Here, we focus on the case where the latter has full rank $m$, as the main interest of the comparison lies in the properties of $W$. Their main result (recast in our notation) states that the matrix $L$ is invertible if and only if 
\begin{equation}
	\label{up}
	\operatorname{rank} \begin{pmatrix} W \\ A \end{pmatrix} = n,
\end{equation}
that is, the $(n+m) \times n$ matrix obtained by stacking $W$ above the constraint matrix $A$ must have full column rank. The equivalence between the conditions
\begin{equation*}
	\operatorname{rank} \begin{pmatrix} W \\ A \end{pmatrix} = n \iff \ker (W) \cap \ker (A) = \{\mathbf{0}\}
\end{equation*}
is immediate to verify, and in the latter we recognize condition~\eqref{wa} examined previously. The necessary and sufficient criterion~\eqref{up} should thus be compared with the results established in Section~3.1.4 ($W$ symmetric). Specifically:

\begin{itemize}
\item 
The condition $\ker (W) \cap \ker (A) = \{\mathbf{0}\}$ is universally necessary for the invertibility of $L$, regardless of any structural properties of $W$ and $A$: indeed, any non-zero vector $\mathbf{x} \in \ker (W) \cap \ker (A)$ yields an element $(\mathbf{x}, \mathbf{0})^T \in \ker L$.
	
\item The condition $\ker (W) \cap \ker (A) = \{\mathbf{0}\}$ is sufficient for the invertibility of $L$ if $W$ is symmetric and sign-semidefinite, in line with our analysis summarized in equation (\ref{cond2wa}).
\end{itemize}
As established in \cite{udwadia2006}, assuming $W$ to be symmetric and positive semidefinite is what renders $\ker (W) \cap \ker (A) = \{\mathbf{0}\}$ a sufficient condition as well (thereby yielding a necessary and sufficient criterion). If $W$ lacks either symmetry or sign-semidefiniteness, the condition remains necessary but ceases to be sufficient.

\noindent
We also note that condition (\ref{cond2}) (along with $A$ having full rank) provides an unconditional invertibility criterion for arbitrary matrices, requiring no structural properties such as symmetry or definiteness.

\subsubsection{The condition $\det(V^T W V) \neq 0$}

\noindent
The non-singularity of the reduced matrix $V^T W V \in \mathbb{R}^{(n-m) \times (n-m)}$ plays a fundamental role in constrained optimization, numerical linear algebra, and the theory of saddle point systems of the KKT form (\ref{Lintr}). In the literature of optimization and linear algebra, $V^T W V$ is commonly known as the Reduced Hessian or Projected Hessian (\cite{cottle,luenberger,rockafellar,wolfe}). 
In modern applications, the condition $\det(V^T W V) \neq 0$ is crucial across several domains, such as saddle point systems for incompressible flows, the numerical analysis of PDEs in Navier-Stokes equations, and Sequential Quadratic Programming (SQP).

\noindent
Proposition~3 establishes that condition (\ref{condeq}), which is necessary and sufficient for the invertibility of $L$, is equivalent to (\ref{cond2}). The matrix $V^T W V$ represents the projection of the operator $W$ onto $\ker(A)$, expressed in the reduced coordinates provided by the basis matrix $V$.

\noindent
In most of the classical literature on numerical linear algebra and optimization (such as \cite{golub,nocedal2006}) the invertibility condition does not typically appear expressed in the exact set-theoretic notation\ (\ref{cond2}). Instead, traditional texts favor the equivalent matrix formulation (\ref{condeq}) which, however, carries the disadvantage of not being coordinate-free, as it requires computing the basis matrix $V$ - an auxiliary object not directly present among the blocks of the given system in (\ref{L}).

\noindent
From a conceptual standpoint, condition~\eqref{cond2} provides an intrinsic, coordinate-free characterization of invertibility directly defined on the underlying subspaces. In contrast, the condition $\det(V^T W V) \neq 0$ offers a computational counterpart that requires constructing an explicit basis matrix $V$ for $\ker(A)$ to evaluate the reduced operator in coordinates.

\section{Conclusion}

\noindent
In this paper, we presented the exact necessary and sufficient algebraic conditions for the invertibility of the saddle-point matrix of type (\ref{Lintr}), which we specifically contextualized for constrained systems subject to equations (\ref{vinc}) and governed by D'Alembert-Lagrange dynamics of the form~\eqref{eqmotolambda}. The invertibility of the corresponding coefficient matrix (\ref{L}) guarantees the possibility of expressing the accelerations $\ddot{\mathbf{q}}$ and the Lagrange multipliers $\bm{\lambda}$ in normal form - that is, explicitly as functions of the Lagrangian state variables $\mathbf{q}$, $\dot{\mathbf{q}}$, and time $t$.
Although in the context of mechanical system dynamics the matrix $W$ - which corresponds to the Hessian of the Lagrangian function, see (\ref{m}) - is positive definite in classic applications (in which case the matrix $L$ is guaranteed to be invertible provided the constraints are independent), we found it of particular interest to investigate the broader question of establishing necessary and sufficient conditions for the invertibility of $L$ in the general case.

\noindent
By establishing that $L$ is non-singular if and only if $\rank(A) = m$ and $\ker(A) \cap \{{\bf x} \in \mathbb{R}^n \mid W{\bf x} \in \mathrm{im}(A^T)\} = \{\mathbf{0}\}$, we provided a necessary and sufficient algebraic condition endowed with a clear geometric meaning, which can be simply summarized as follows: the mapping $W$ must not transform any non-zero vector $\mathbf{x} \in \ker(A)$ into a vector orthogonal to the entire kernel of $A$. 

\noindent
The two necessary and sufficient conditions for the invertibility of $L$ find a physical interpretation based on the duality between constraint kinematics and system inertia. In the specific context of constrained system dynamics, the intersection condition plays the role of a dynamic non-singularity requirement. Physically, it prevents unconstrained massless motions by ensuring that any zero-mass direction ($W\mathbf{x} = \mathbf{0}$) is fully suppressed by constraints and does not lie in $\ker(A)$, which would otherwise leave accelerations along such directions indeterminate. Furthermore, it guarantees effective inertia along virtual displacements: any non-zero virtual acceleration $\mathbf{x} \in \ker(A)$ generates inertial forces $W\mathbf{x}$ that cannot be entirely balanced by constraint reactions $\operatorname{im}(A^T)$, thereby ensuring a uniquely determined dynamic response. The action of $W$ on ${\bf x}$ must be ``energetically active'' along the directions left unconstrained by $A$.

\noindent
The new formal interpretation of the invertibility conditions (\ref{cond1}) and (\ref{cond2}) provides an opportunity to compare them with analogous conditions from classical results: we can summarize the analysis carried out in Section 3.1 through the following scheme, where "nec." or "suff." denote a necessary or sufficient condition for the invertibility of $L$ (assuming that the necessary condition (\ref{cond1}) $\rank(A)=m$ always holds):

$$
\begin{array}{l}
	(1) \;\; W\;\textrm{positive definite}\,(suff.)\;\Longrightarrow\;
	{\bf x}^TW{\bf x}\not =0\;\forall {\bf x}\in A\setminus\{ {\bf 0}\} \,(suff.)\\
	\Longrightarrow\;\ker(A) \cap \{{\bf x} \mid W{\bf x} \in \mathrm{im}(A^T)\} = \{\mathbf{0}\}\;(nec.\,\&\,suff.)\\
	\\
	(2)\; \; \det(V^TWV)\not =0 \;\Longleftrightarrow\;
	\ker(A) \cap \{{\bf x} \mid W{\bf x} \in \mathrm{im}(A^T)\} = \{\mathbf{0}\}\;(both\;nec.\,\&\,suff.)\\
	\\
	(3)\; \; W\;\;\textrm{symmetric and positive semidefinite}: \ker(A)\cap \ker(W)=\{ {\bf 0}\}\\ \Longleftrightarrow\;
	\ker(A) \cap \{{\bf x}\mid W{\bf x} \in \mathrm{im}(A^T)\} = \{\mathbf{0}\}\;(both\;nec.\,\&\,suff.)
\end{array}
$$
The preceding statements are stated, proved, and developed in Section 3.1. In addition to our condition (\ref{cond2}), the conditions appearing therein refer to classical results, such as globally positive-definite mass matrices, kernel ellipticity requirements, and symmetric $W$. In Section 3.2, we briefly introduced the methods to contextualize them within the conditions discussed and to establish a comparison with our conditions stated in Proposition 1 (main result).

\noindent
We highlight what we consider to be the primary advantages of formulation (\ref{cond2}) as an invertibility condition.
Referring to the classification above, item $(1)$ illustrates that the proposed condition $\ker(A) \cap \{{\bf x} \mid W{\bf x} \in \mathrm{im}(A^T)\} = \{\mathbf{0}\}$ fully characterizes invertibility, unlike standard assumptions of sign definiteness or kernel ellipticity, which serve only as sufficient conditions. The domain of applicability is thus maximally extended, yielding the minimal necessary geometric requirement to prevent a non-trivial kernel in the overall system. Furthermore, there exist dynamical systems where the matrix $W$ is not positive definite - such as non-standard Lagrangians or models originating outside classical mechanics - and our criterion addresses these settings directly.
Regarding point $(2)$, our condition does not require computing a basis $V$ for $\ker(A)$ (and is therefore coordinate-free). Finally, while criterion $(3)$ is both necessary and sufficient, it applies only to a restricted class of matrices $W$, whereas our criterion holds universally without requiring $W$ to be symmetric or positive semidefinite.
Furthermore, by avoiding the computation of Moore--Penrose pseudoinverses, the proposed approach yields a direct, coordinate-free formulation that avoids potential numerical instabilities associated with rank-revealing matrix factorizations.

\noindent
Several promising avenues for future investigation naturally emerge from this work. The first crucial step involves generalizing the criterion to rank-deficient constraint matrices ($\mathrm{rank}(A) < m$), extending this exact necessary and sufficient framework to systems featuring kinematically redundant or singular constraints, where $A$ fails to have full row rank. While classical approaches - most notably Udwadia's fundamental equation of motion - handle rank deficiency via projection operators and generalized inverses ($A^+$), we aim to formulate an intrinsic algebraic characterization of solvability and consistency that avoids pseudoinverses even when constraints are linearly dependent.
Furthermore, another compelling extension consists in relaxing the assumption of ideal/smooth constraints - thus departing from the strict validity of (\ref{spostvirt}) - by replacing the zero right-hand side with a general function of $\mathbf{q}$, $\dot{\mathbf{q}}$, and $t$.
The linear-algebraic framework adopted here promises to remain effective when incorporating these extensions.

\end{document}